\documentclass[%
reprint,
superscriptaddress,
 amsmath,amssymb,
 aps,
]{revtex4-1}

\usepackage[T1]{fontenc} 

\usepackage{graphicx}
\usepackage{dcolumn}
\usepackage{bm}
\usepackage[ngerman]{babel}

\usepackage{lipsum}
\usepackage{color}
\usepackage[normalem]{ulem}
\usepackage{simplewick}
\usepackage{qcircuit}
\usepackage{braket}
\usepackage{float}
\usepackage{multirow}
\usepackage{tikz}
\usepackage{tabu}
\usepackage{xr-hyper}
\usepackage[colorlinks=true,urlcolor=blue,citecolor=blue,linkcolor=blue]{hyperref}
\usepackage{subcaption}
\usepackage{multirow}
\usepackage{ragged2e}
\usepackage{etoolbox}
\usepackage{booktabs}
\usepackage{bbm}
\usepackage{mathtools}
\usepackage{url}
\usepackage{lipsum}

\usepackage{array}
\newcolumntype{L}[1]{>{\raggedright\let\newline\\\arraybackslash\hspace{0pt}}m{#1}}
\newcolumntype{C}[1]{>{\centering\let\newline\\\arraybackslash\hspace{0pt}}m{#1}}
\newcolumntype{R}[1]{>{\raggedleft\let\newline\\\arraybackslash\hspace{0pt}}m{#1}}

\usepackage[section]{placeins}

\newcommand{\minus}{\scalebox{0.75}[1.0]{$-$}}

\definecolor{fifiColor}{HTML}{03ad29}

\begin{document}

\title{Quanteninformationsverarbeitung in der Gymnasialen Oberstufe}% \\

\author{Andreas J. C. Woitzik}
\email{andreas.woitzik@physik.uni-freiburg.de}
\affiliation{Physikalisches Institut, Albert-Ludwigs-Universit\"at Freiburg, Hermann-Herder-Stra\ss e 3, D-79104 Freiburg im Breisgau, Bundesrepublik Deutschland}

\date{\today}

\begin{abstract}\noindent
Die Quanteninformationsverarbeitung wird üblicherweise in den höheren Semestern des Physikstudiums als Wahlbereich gelehrt. 
Wir zeigen, dass ein grundlegendes Verständnis der Quanteninformation in der Gymnasialen Oberstufe möglich ist.
Darüber hinaus argumentieren wir, warum die Behandlung der Quanteninformation zu diesem Zeitpunkt sinnvoll ist und beschreiben eine Unterrichtseinheit, die bereits in der Praxis erprobt wurde. 
Für die Umsetzung der Einheit gehen wir auf konkrete didaktische Reduktionen ein und beschreiben kurz die benötigten fachlichen Grundlagen.
\end{abstract}

\pacs{Valid PACS appear here}
\maketitle
\

%%%%%%%%%%%%%%%%%%%%%%%%%%%%%%%%%%%%%%%%%%%%%%%%%%%%%
\section{Einleitung und Motivation}
%%%%%%%%%%%%%%%%%%%%%%%%%%%%%%%%%%%%%%%%%%%%%%%%%%%%%
Die Quanteninformationsverarbeitung ist ein verhältnismäßig junger Forschungszweig, dem sowohl im wissenschaftlichen Kontext als auch in den Medien ein großes Potenzial zugeschrieben wird und dementsprechend viel Beachtung geschenkt wird.
Die Anwendungen reichen von der Kryptographie über die Simulation von Molekülen hin zur Metrologie und nutzen alle Besonderheiten der Quanteninformation, auf die wir im Folgenden teilweise eingehen werden.\\

Die Boolsche Algebra ist für uns heute der gewohnte Repräsentant für Information zur Datenverarbeitung. 
Doch sie ist nicht die einzige Darstellung von Information und Informatikunterricht sollte ein Verständnis dafür geben, was die Vor- und Nachteile verschiedener Repräsentationsformen von Information sind.
Eine gute Möglichkeit, dieses Verständnis zu schärfen, bietet die Behandlung der Quanteninformationsverarbeitung im Oberstufenunterricht. \\

Drei grundlegende Bausteine der Informationsverarbeitung sind (a) der Informationsträger, (b) die Operationen, die mit ihm unternommen werden können und (c) die Extraktion von Information.
Wir werden diese drei Grundbausteine für die Quanteninformationsverarbeitung im Unterrichtsgeschehen betrachten und ihre Unterschiede zur digital-binären Informationsverarbeitung aufzeigen.
So lassen sich verblüffende Unterschiede feststellen: die Unmöglichkeit des allgemeinen Kopierens von Quanteninformation, die invasive Messung oder auch die Reversibilität aller Berechnungen.
Diese Unterschiede brechen mit einigen Selbstverständnissen der klassischen Informationsverarbeitung.
Hieraus ergeben sich wiederum interessante Konsequenzen wie zum Beispiel ein Schlüsselaustauschprotokoll (siehe Kapitel~\ref{sec:bb84}). \\

Es gibt bis dato verhältnismäßig wenige Arbeiten zur Quanteninformationsverarbeitung in der Schule und diese nehmen einen physikalischen Blickwinkel ein~\cite{kublbeck2002, dur2010, dur2012, kohnle2016}.
Ebenso verhält es sich mit Lehrbüchern der Hochschule, sowohl im Englischen als auch im Deutschen~\cite{Nielsen_Chuang2011, brands2011}. 
Eine neuere Arbeit beschäftigt sich mit den Schülervorstellungen zur Quanteninformationsverarbeitung~\cite{schorn2019}. 
In dieser wird festgestellt, dass selbst bei einer Selektion besonders leistungsstarker Schüler viele Fehlvorstellungen zur Quanteninformation bestehen.
Außerdem werden hier bereits teilweise ähnliche didaktische Reduktionen wie unsere vorgeschlagen, allerdings wieder vor einem physikalischen Hintergrund. 
Die vorliegende Arbeit beschäftigt sich mit der Quanteninformationsverarbeitung aus informatischer Perspektive und versucht bewusst, physikalische Parallelen auszusparen, um den Blick auf das aus informationstheoretischer Sicht Wesentliche zu lenken.\\

Wir beschreiben mögliche fachdidaktische Reduktionen für eine Unterrichtseinheit in der gymnasialen Oberstufe, die ein grundlegendes Verständnis der (mathematischen) Besonderheiten der Quanteninformation schaffen soll. 
Der vorliegende Text richtet sich an Unterrichtspraktiker der Informatik mit soliden mathematischen Grundlagen sowie an Fachdidaktiker aus dem universitären Kontext.
Die entsprechenden Unterrichtsmaterialen werden auf Anfrage kostenlos zur Verfügung gestellt.

%%%%%%%%%%%%%%%%%%%%%%%%%%%%%%%%%%%%%%%%%%%%%%%%%%%%%
\section{Fachliche Einordnung}
%%%%%%%%%%%%%%%%%%%%%%%%%%%%%%%%%%%%%%%%%%%%%%%%%%%%%

Die Quanteninformationsverarbeitung ist kein Thema der Informatikbildungspläne der Bundesländer.
Sie wird klassischerweise in den fortgeschrittenen Jahren des Physikstudiums, seltener auch des Informatikstudiums, gelehrt. 
Da die Quanteninformationsverarbeitung vornehmlich von Physikern erforscht wird, setzt die Literatur sehr häufig ein grundlegendes Verständnis der Physik - insbesondere der Quantenmechanik - voraus. 
Dabei ist die Behandlung der Materie, reduziert auf die informationstheoretischen Aspekte, für die Informatik in der Schule durchaus geeignet. 
Die Schüler benötigen lediglich Grundlagen der analytischen Geometrie in reellen Vektorräumen sowie der Boolschen Algebra und Operationen darauf, die durch Gatter dargestellt werden können.

%%%%%%%%%%%%%%%%%%%%%%%%%%%%%%%%%%%%%%%%%%%%%%%%%%%%%
\section{Rahmenbedingungen}
%%%%%%%%%%%%%%%%%%%%%%%%%%%%%%%%%%%%%%%%%%%%%%%%%%%%%

Die vorliegende Unterrichtseinheit wurde für einen vierstündigen Informatikkurs entwickelt und im Jahr 2018 am Faust Gymnasium in Staufen im Breisgau gehalten. Zeitpunkt der Einheit war nach dem schriftlichen und vor dem mündlichen Abitur. Die Größe der Klasse betrug zwölf Schüler und es wurden vier aufeinanderfolgende Doppelstunden unterrichtet. Es fand keine Überprüfung des Lernerfolgs der Einheit statt. 

%%%%%%%%%%%%%%%%%%%%%%%%%%%%%%%%%%%%%%%%%%%%%%%%%%%%%
\section{Lernziele}\label{sec:Lernziele}
%%%%%%%%%%%%%%%%%%%%%%%%%%%%%%%%%%%%%%%%%%%%%%%%%%%%%

Es soll ein grundlegendes Verständnis für die Quanteninformationsverarbeitung in vereinfachter Form geschaffen werden.

\begin{itemize}
\item Die Schüler können Quantenbits (QuBits) auf einem Einheitskreis darstellen und die Zustände, die klassischen Bits entsprechen, benennen.
\item Die Schüler können einfache Gatter (X, Z, H und CX) auf gegebene QuBits anwenden.
\item Die Schüler können das BB84-Protokoll anwenden und erklären, was die Stärken und Schwächen der Quanteninformation für den Informationsaustausch sind.
\item Die Schüler können die Quantenteleportation beschreiben. 
\item Die Schüler können verschiedene Realisierungen von QuBits in der Natur und Technik nennen.
\end{itemize}

%%%%%%%%%%%%%%%%%%%%%%%%%%%%%%%%%%%%%%%%%%%%%%%%%%%%%
\section{Einstieg in die Einheit}\label{sec:Einstieg}
%%%%%%%%%%%%%%%%%%%%%%%%%%%%%%%%%%%%%%%%%%%%%%%%%%%%%

\begin{figure*}[htb!]
  \centering
\includegraphics[width = 0.7\textwidth]{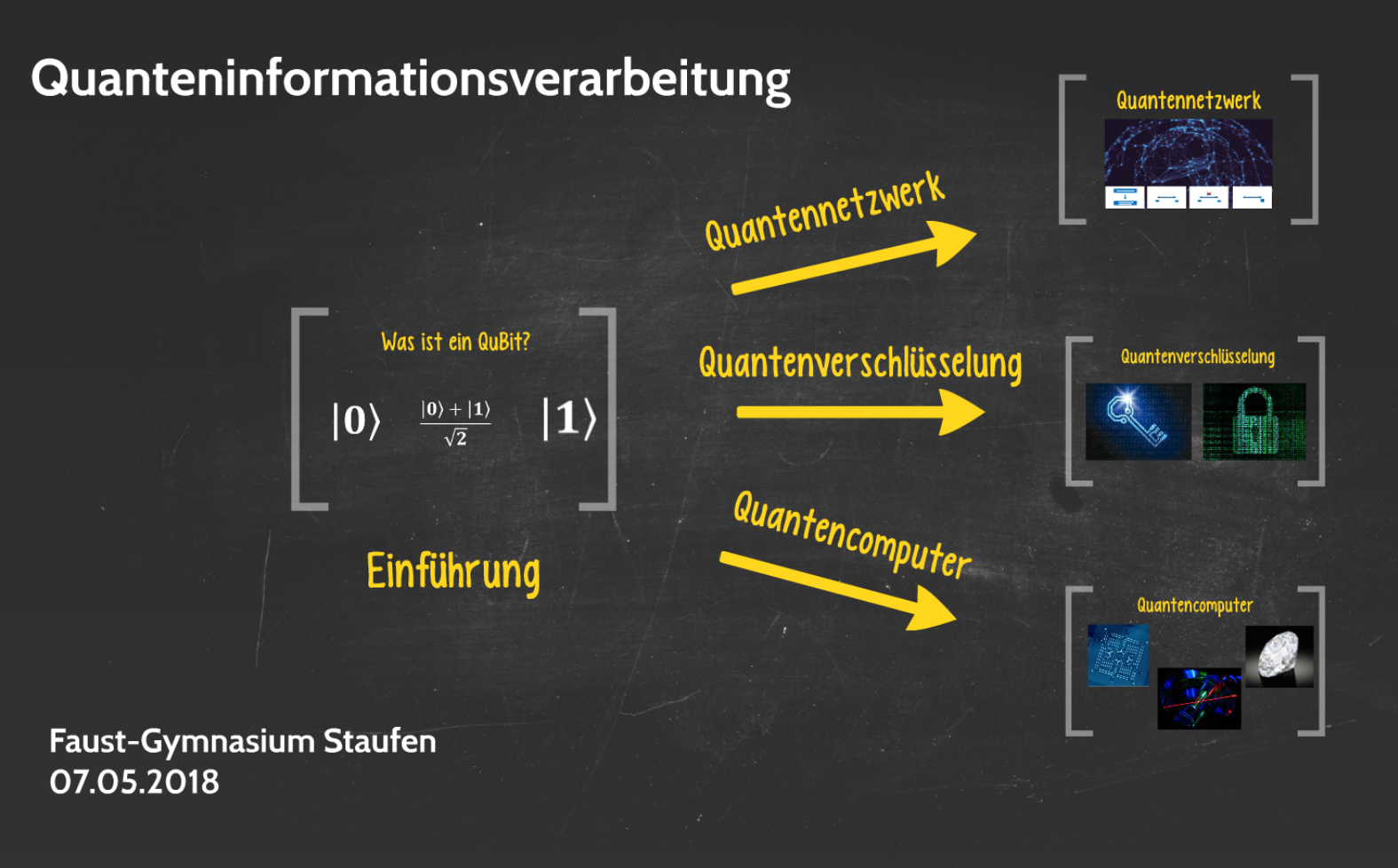}
\caption{Screenshot der Einstiegspräsentation. Es wurden die Gebiete Quantennetzwerke, Quantenverschlüsselung und Quantencomputer angesprochen.}
\label{fig:prezi}
\end{figure*}

Zu Beginn der Einheit wurde eine Sammlung an Vorwissen und Erwartungen zusammengestellt. Die Erwartungen sind in die Planung der Doppelstunden zwei bis vier eingeflossen. 
Der Abgleich der Erwartungen erfolgte mit einer kurzen Präsentation mit einem Überblick zu den gängigen Themen der Quanteninformationsverarbeitung~(siehe Abbildung \ref{fig:prezi} im Anhang).

%%%%%%%%%%%%%%%%%%%%%%%%%%%%%%%%%%%%%%%%%%%%%%%%%%%%%
\section{Aufbau der Unterrichtseinheit}\label{sec:AufbauDerEinheit}
%%%%%%%%%%%%%%%%%%%%%%%%%%%%%%%%%%%%%%%%%%%%%%%%%%%%%

Der Unterrichtsverlauf entspricht im Groben der Einführung in die Quanteninformationsverarbeitung, wie sie auch im universitären Kontext gelehrt wird. 
Das Standardwerk der Literatur von Nielsen und Chuang, \textit{Quantum Computation and Quantum Information}~\cite{Nielsen_Chuang2011}, beginnt ebenfalls mit der Einführung der mathematischen Struktur des Quantenbits (QuBit) und lässt Gatteroperationen und einfache Algorithmen folgen.
Durch die Unabhängigkeit der Themen wie Quantennetzwerke, Quantenverschlüsselung und der Hardware der Quantencomputer ist der Aufbau der Einheit variabel. 
Ebenso ist es gut denkbar, die Einheit auf zwei Doppelstunden zu kondensieren oder weiter auszubauen, zum Beispiel für Projekttage.

\subsection{Das Quantenbit (QuBit)}

\begin{figure}[H]
\begin{center}
\begin{tikzpicture}
\node[inner sep=0pt] (russel) at (2,0)
    {\includegraphics[width=0.7\linewidth]{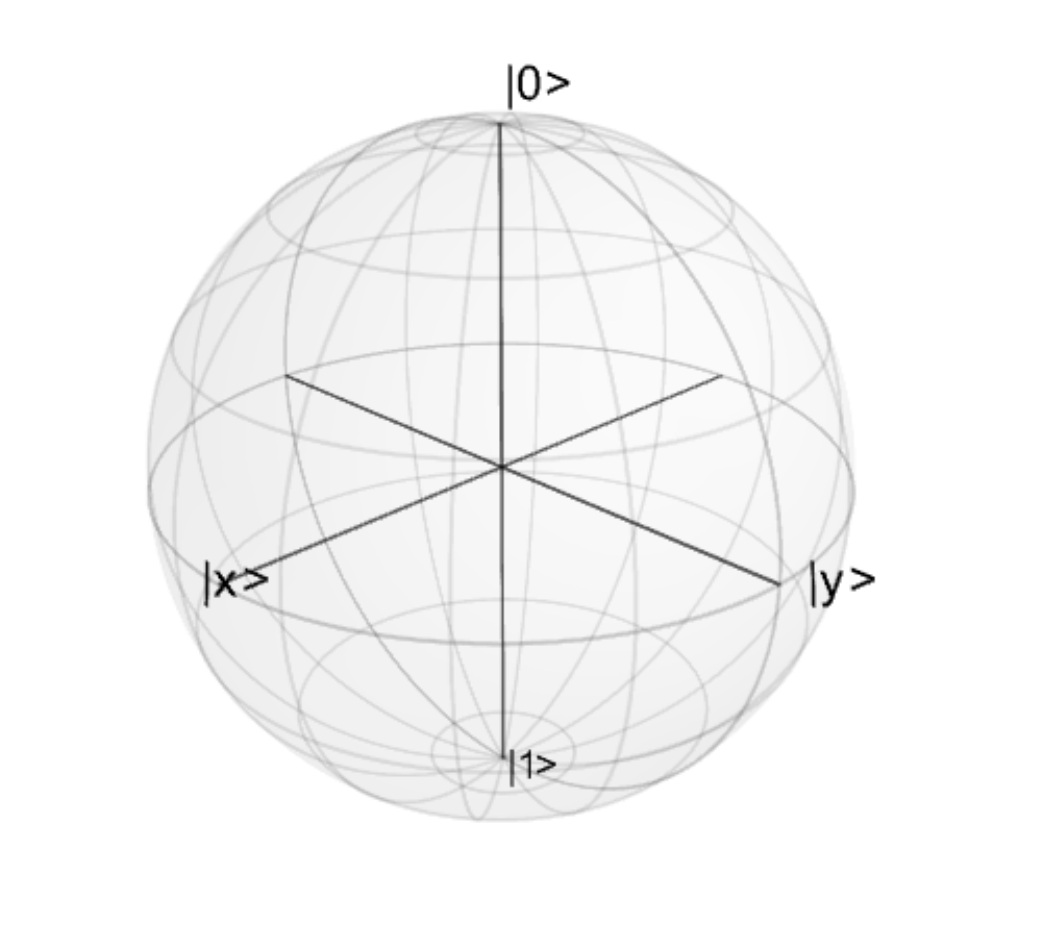}};
\end{tikzpicture}
\end{center}
\caption{Die Bloch-Kugel ist das Standardmodell zur Veranschaulichung eines Quantenzustands, der durch ein QuBit repräsentiert werden kann. Der Nachteil des Modells für die Schule ist, dass orthogonale Zustände im Hilbertraum auf der Bloch-Kugel Antipoden bilden.}
\label{fig:blochkugel} 
\end{figure}

Die zentrale didaktische Reduktion der Einheit geschieht mit der Einführung des Quantenbits. Ein QuBit $\ket{q}$ wird im akadamischen Kontext definiert als normiertes Element eines zweidimensionalen komplexen Vektorraums $\mathcal{H}$ mit Skalarprodukt (ein Hilbertraum), für das eine globale Phase $\exp(i \phi)$ keine Bedeutung hat~\footnote{Wir verwenden die von Dirac eingeführt BraKet-Schreibweise, auch wenn wir ihre mathematische Eleganz nicht nutzen, sondern damit lediglich ausdrücken wollen, dass es sich bei einem Objekt um ein QuBit handelt.}. 
Eine mögliche Veranschaulichung der Zustände eines QuBits ist die Bloch-Kugel~(siehe Abbildung~\ref{fig:blochkugel}), auf der orthogonale Zustände Antipoden bilden.
Dieses Bild ist in der Physik weit verbreitet, da es auch zur Beschreibung des Polarisationszustandes eines Photons verwendet werden kann.

\begin{figure}[H]
\begin{center}
\begin{tikzpicture}
\node[inner sep=0pt] (russel) at (2,0)
    {\includegraphics[width=0.5\linewidth]{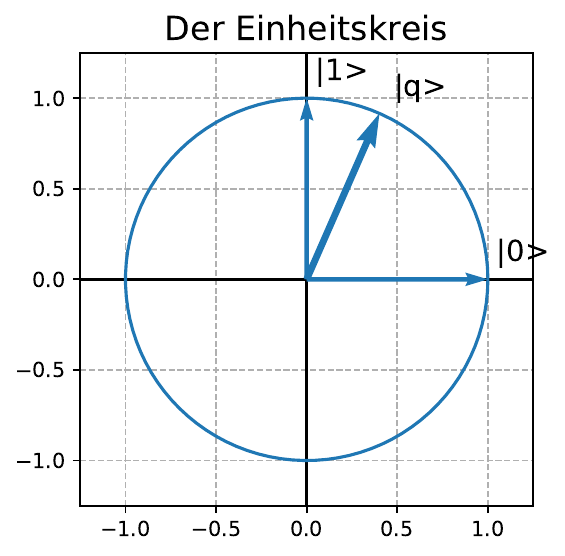}};
\end{tikzpicture}
\end{center}
\caption{Der Einheitskreis zur Veranschaulichung eines QuBits $\ket{q}$, das durch einen Vektor beschrieben wird.}
\label{fig:einheitskreis} 
\end{figure}

Für Schüler ist es allerdings verwirrend, dass orthogonale Zustände auf der Kugel Antipoden bilden, also linear abhängig erscheinen. 
Außerdem reicht für die zentralen Phänomene der Quanteninformationsverarbeitung eine Betrachtung des QuBits als Element der 1-Sphäre (des Einheitskreises) aus.
Wir schlagen deshalb folgende Definition eines QuBits für den Informatikunterricht vor:

Ein QuBit wird als Vektor eines Elements des Einheitskreises beschrieben (siehe Abbildung~\ref{fig:einheitskreis}).
Seien dafür $\alpha, \beta \in \mathbb{R}$, dann ist ein QuBit $\ket{q}$ definiert als:
\begin{align*}
\ket{q} = \begin{pmatrix}
\alpha \\
\beta
\end{pmatrix} \quad \text{mit} \quad \alpha^2 + \beta^2 =  1 .
\end{align*}

Außerdem definieren wir die Zustände, die dem 0-Zustand beziehungsweise dem 1-Zustand eines klassischen Bits entsprechen als:
\begin{align*}
0_B = \ket{0} =  \begin{pmatrix}
1 \\
0
\end{pmatrix} \quad \text{und} \quad 1_B = \ket{1} = \begin{pmatrix}
0 \\
1
\end{pmatrix} . 
\end{align*}

\subsection{Ein-QuBit-Quantengatter}
\begin{figure*}[htb!]
  \centering
\includegraphics[width = 0.24\textwidth]{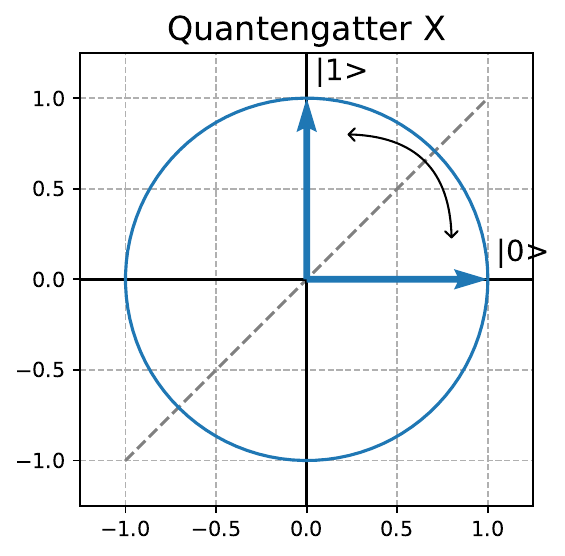}
\includegraphics[width = 0.24\textwidth]{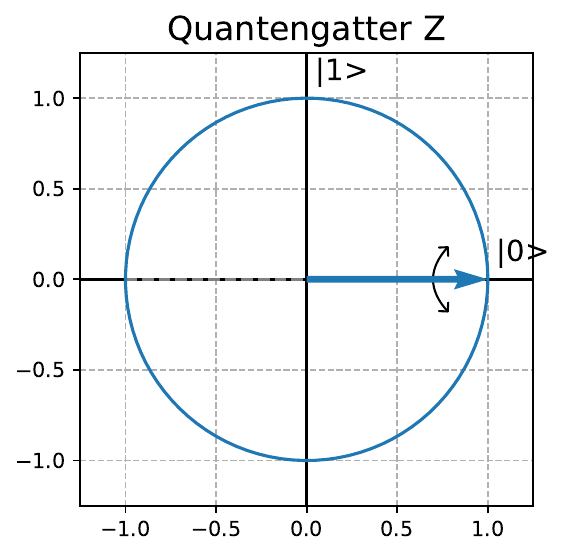}
\includegraphics[width = 0.24\textwidth]{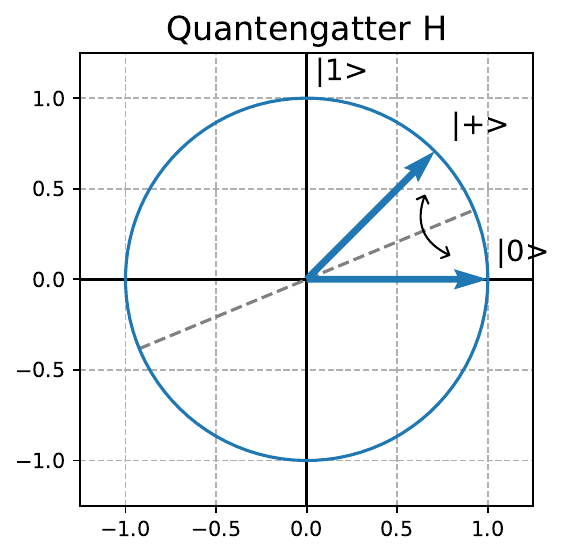}
\includegraphics[width = 0.24\textwidth]{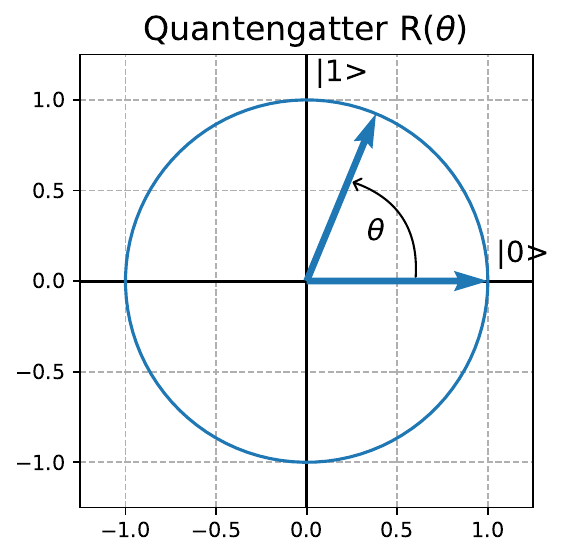}
\caption{Die drei vorgestellten Quantengatter X, Z und H beschreiben alle Spiegelungen um eine definierte Achse. Daneben gibt es die Möglichkeit der Rotation $R(\theta)$ um einen beliebigen Winkel $\theta$.}
\label{fig:gatter1qubit}
\end{figure*}

In der Quanteninformationsverarbeitung werden Gatter durch unitäre Matrizen beschrieben.
Diese beschreiben Drehspiegelungen in hochdimensionalen Hilberträumen.
Entsprechend können wir unsere Ein-QuBit-Gatter als Drehspiegelungen am Einheitskreis deuten; dadurch erklärt sich die Reversibilität der Gatter sofort. 
Neben der geometrischen Sichtweise wurden die Gatter in Form von Wahrheitstabellen mit den Basiszuständen eingeführt (siehe Tabelle~\ref{tab:1qubitgatter}). 
Somit kann direkt an Bekanntes aus der (klassischen) Informatik angeschlossen werden.
Es können theoretisch sämtliche Drehspiegelungen auf dem Einheitskreis realisiert werden. Für die Unterrichtseinheit reichen allerdings die ($X, Z ,H$)-Gatter, die ein Pauli-X, Pauli-Z und ein Hadamard-Gatter realisieren.
Wir geben hier dennoch auch eine Verbildlichung der Drehung $R(\theta)$ an, da sie für Schüler konzeptionell kein Problem darstellt und wir somit der Vollständigkeit genüge tun.
Das (Pauli-)X entspricht in unserem Bild einer Spiegelung an der ersten Winkelhalbierenden, das \text{(Pauli-)Z} an der horizontalen Achse und das H-Gatter (für Hadamard) einer Spiegelung an der $22,5^\circ$-Achse, wobei sich der Winkel auf die x-Achse bezieht (siehe Abbildung~\ref{fig:gatter1qubit}).

\begin{table}[htb]
\setlength{\tabcolsep}{3.5pt}
\[
\begin{array}{c | c}
  \ket{q} & X\ket{q}  \\ \hline 
   \ket{0} & \ket{1} \\
     \ket{1} & \ket{0} \\
\end{array} 
\hspace{0.5cm}
\begin{array}{c | c}
  \ket{q} & Z\ket{q}  \\ \hline 
   \ket{0} & \ket{0} \\
     \ket{1} &  \minus\ket{1} \\
\end{array}
\hspace{0.5cm}
\begin{array}{c | c}
  \ket{q} & H\ket{q}  \\ \hline 
   \ket{0} & \ket{+} \\
     \ket{1} & \ket{-} \\
\end{array}\]
\[
\begin{array}{c | c}
  \ket{q} & R(\theta)\ket{q}  \\ \hline 
   \ket{0} & \cos(\theta)\ket{0}+\sin(\theta)\ket{1} \\
     \ket{1} & -\sin(\theta)\ket{0}+\cos(\theta)\ket{1}
\end{array}
\]
\caption{Wahrheitstabellen der 1-QuBit-Gatter: $X, Z, H$ und $R(\theta)$}\label{tab:1qubitgatter}
\end{table}

\subsection{Die Messung}
\begin{figure}[H]
\begin{center}
\begin{subfigure}{0.20\textwidth}
   \includegraphics[width=1\textwidth]{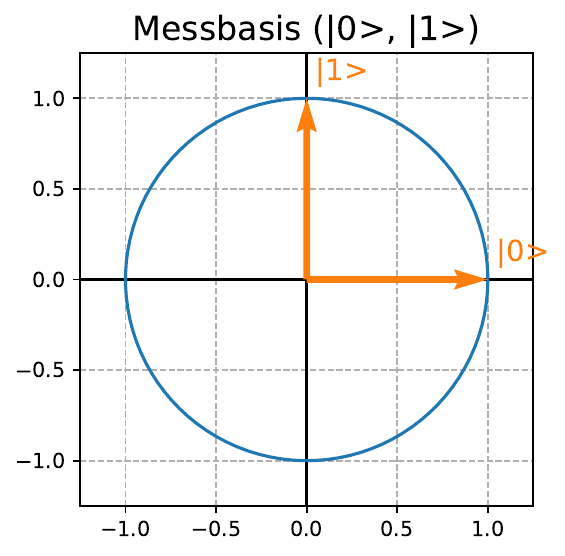}
      \caption[]{}
      \label{fig:I1}
  \end{subfigure}%
\begin{subfigure}{0.20\textwidth}
   \includegraphics[width=1\textwidth]{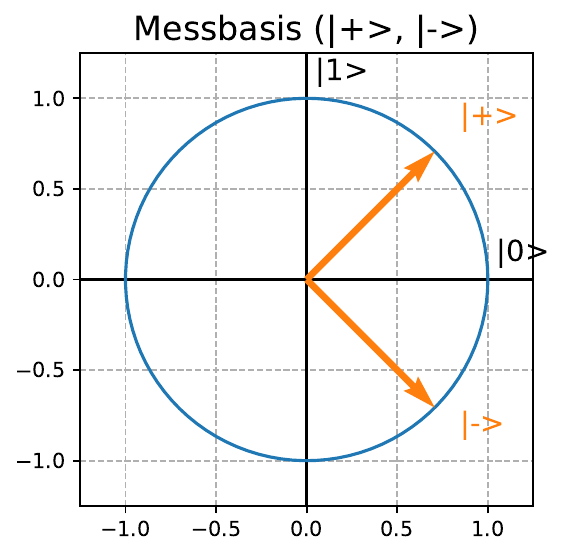}
      \caption[]{}
      \label{fig:I1}
  \end{subfigure}
\end{center}
\caption{Gezeigt sind die Messbasen (a) $(\ket{0},\ket{1})$ und (b) $(\ket{+},\ket{-})$.}
\label{fig:messung} 
\end{figure}

Grundlegend verschieden zur klassischen Informatik ist das Auslesen von Information, sprich die Messung eines Quantenzustands.
Da sie diesen verändert, sprechen wir von einer invasiven Messung.
Für eine Messung muss zunächst eine Messbasis gewählt werden. Das ist eine orthogonale Basis des Vektorraums des Zustands.
Die Wahrscheinlichkeit für ein bestimmtes Messergebnis ist hierbei durch das Quadrat des Skalarprodukt des Zustands (vor der Messung) mit dem entsprechenden Basisvektor gegeben.
Diese Projektion ist ein unstetiger und irreversibler Vorgang, im Kontrast zu den stets reversiblen Quantengattern.
Wir benötigen für das Verständnis und später das BB84-Protokoll lediglich die Messbasen $(\ket{0}, \ket{1})$ und $(\ket{+}, \ket{-})$, die in Abbildung~\ref{fig:messung} veranschaulicht werden. 

Die Veränderung des Zustands durch die Messung bildet die Grundlage für das hier vorgestellte Verschlüsselungsprotokoll BB84.

\subsection{Quantenverschlüsselung}
\label{sec:bb84}
Das behandelte Protokoll BB84 ist ein Schlüsselaustauschprotokoll. Es wurde im Jahr 1984 von Bennett und Brassard vorgeschlagen~\cite{bennett14}.
Im Gegensatz zu vielen modernen Quantenprotokollen zum Schlüsselaustausch verwendet das Protokoll keine verschränkten Zustände und ist leicht verständlich.
Außerdem wird durch das Protokoll besonders schön deutlich, dass das No-Cloning-Theorem und die invasive Messung wichtig für die Quantenkryptographie sind.

\begin{enumerate}
\item Alice erzeugt zufällige Bits durch Münzwurf.
\begin{center}
\begin{tikzpicture}
\draw[step=0.5cm,color=gray] (-1,0.5) grid (3,1);
\node at (-.75,+0.75) {0};
\node at (-0.25,+0.75) {1};
\node at (+0.25,+0.75) {1};
\node at (+0.75,+0.75) {1};
\node at (1.25,+0.75) {0};
\node at (1.75,+0.75) {0};
\node at (+2.25,+0.75) {1};
\node at (+2.75,+0.75) {0};
\end{tikzpicture}
\end{center}

\item Alice entscheidet sich zufällig durch Münzwurf für eine Messrichtung je QuBit.
\begin{center}
\begin{tikzpicture}
\draw[step=0.5cm,color=gray] (-1,0.5) grid (3,1);
\node at (-.75,+0.75) {+};
\node at (-0.25,+0.75) {+};
\node at (+0.25,+0.75) {x};
\node at (+0.75,+0.75) {x};
\node at (1.25,+0.75) {+};
\node at (1.75,+0.75) {x};
\node at (+2.25,+0.75) {+};
\node at (+2.75,+0.75) {+};
\end{tikzpicture}
\end{center}

\item Wenn Alice beim Münzwurf der Messrichtung eine 0 erzeugt hat, schickt sie ein QuBit entlang der Richtung $(\ket{0}, \ket{+})$ oder bei 1 entlang der jeweils orthogonalen Richtung $(\ket{1}, \ket{-})$.
\begin{center}
\begin{tikzpicture}
\draw[step=0.5cm,color=gray] (-1,0.5) grid (3,1);
\node at (-.75,+0.75) {$\ket{0}$};
\node at (-0.25,+0.75) {$\ket{1}$};
\node at (+0.25,+0.75) {$\ket{-}$};
\node at (+0.75,+0.75) {$\ket{-}$};
\node at (1.25,+0.75) {$\ket{0}$};
\node at (1.75,+0.75) {$\ket{+}$};
\node at (+2.25,+0.75) {$\ket{1}$};
\node at (+2.75,+0.75) {$\ket{0}$};
\end{tikzpicture}
\end{center}

\item Alice sendet die QuBits an Bob. 

\item Bob entscheidet sich zufällig durch Münzwurf für eine Messrichtung für jedes QuBit und misst die QuBits.
\begin{center}
\begin{tikzpicture}
\draw[step=0.5cm,color=gray] (-1,0.5) grid (3,1.5);
\node at (-.75,+0.75)  {$\ket{0}$};
\node at (-0.25,+0.75) {$\ket{+}$};
\node at (+0.25,+0.75) {$\ket{-}$};
\node at (+0.75,+0.75) {$\ket{1}$};
\node at (1.25,+0.75) {$\ket{0}$};
\node at (1.75,+0.75) {$\ket{+}$};
\node at (+2.25,+0.75) {$\ket{-}$};
\node at (+2.75,+0.75) {$\ket{0}$};
\node at (-.75,+1.25) {+};
\node at (-0.25,+1.25) {x};
\node at (+0.25,+1.25) {x};
\node at (+0.75,+1.25) {+};
\node at (1.25,+1.25) {+};
\node at (1.75,+1.25) {x};
\node at (+2.25,+1.25) {x};
\node at (+2.75,+1.25) {+};
\end{tikzpicture}
\end{center}

\item Bob übersetzt QuBits in Richtung $(\ket{0}, \ket{+})$ zu 0 und in der jeweils orthogonalen Richtung $(\ket{1}, \ket{-})$ zu 1.  
\begin{center}
\begin{tikzpicture}
\draw[step=0.5cm,color=gray] (-1,0.5) grid (3,1.5);
\node at (-.75,+0.75) {$0$};
\node at (-0.25,+0.75) {$0$};
\node at (+0.25,+0.75) {$1$};
\node at (+0.75,+0.75) {$1$};
\node at (1.25,+0.75) {$0$};
\node at (1.75,+0.75) {$0$};
\node at (+2.25,+0.75) {$1$};
\node at (+2.75,+0.75) {$0$};
\node at (-.75,+1.25) {+};
\node at (-0.25,+1.25) {x};
\node at (+0.25,+1.25) {x};
\node at (+0.75,+1.25) {+};
\node at (1.25,+1.25) {+};
\node at (1.75,+1.25) {x};
\node at (+2.25,+1.25) {x};
\node at (+2.75,+1.25) {+};
\end{tikzpicture}
\end{center}

\item Alice \& Bob tauschen öffentlich aus, welche Messrichtungen sie verwendet haben und verwerfen alle Bits, bei denen sie unterschiedliche Messrichtungen verwendet haben.
\begin{center}
\begin{tikzpicture}
\draw[step=0.5cm,color=black] (-1,0.5) grid (3,2.);

\node at (-.75,+1.25) {+};
\node at (-0.25,+1.25) {x};
\node at (+0.25,+1.25) {x};
\node at (+0.75,+1.25) {+};
\node at (1.25,+1.25) {+};
\node at (1.75,+1.25) {x};
\node at (+2.25,+1.25) {x};
\node at (+2.75,+1.25) {+};

\node at (-.75,+.75) [fill=green!20] {0};
\node at (-0.25,+.75) {};
\node at (+0.25,+.75) [fill=green!20] {1};
\node at (+0.75,+.75) {};
\node at (1.25,+.75) [fill=green!20] {0};
\node at (1.75,+.75) [fill=green!20] {0};
\node at (+2.25,+.75) {};
\node at (+2.75,+.75) [fill=green!20] {0};

\node at (-.75,+1.75) {+};
\node at (-0.25,+1.75) {+};
\node at (+0.25,+1.75) {x};
\node at (+0.75,+1.75) {x};
\node at (1.25,+1.75) {+};
\node at (1.75,+1.75) {x};
\node at (+2.25,+1.75) {+};
\node at (+2.75,+1.75) {+};
\end{tikzpicture}
\end{center}

\item Die übrigen Bits sind der Schlüssel und können beispielsweise für eine One-Time-Pad-Verschlüsselung genutzt werden. 
\end{enumerate}

Mit einem teilweisen Abgleich der Schlüssel können Alice und Bob nun auch überprüfen, ob ihr Kanal abgehört wurde.
Falls ein Abhörer (meist Eve genannt) die QuBits von Alice abfängt und ausliest, wählt er mit $50\%$ Wahrscheinlichkeit eine andere Messrichtung als Alice.
In diesem Fall beträgt die Wahrscheinlichkeit, dass Alice und Bob die gleichen Schlüsselbits haben, nur noch $50\%$. 
Sie können durch den Abgleich einiger Bits mit hoher Wahrscheinlichkeit herausfinden, ob jemand mit dem Kanal interagiert.
Das No-Cloning-Theorem ist die theoretische Basis, weshalb diese Überprüfung der Datenintegrität funktioniert.
Würde es nicht gelten, könnte Eve Kopien der QuBits machen und so ungemerkt den Schlüssel erhalten, falls sie die Kopien in den zwei verschiedenen Messbasen misst und die Ergebnisse speichert, bis Alice und Bob ihre Basen kommunizieren.

\subsection{Verschränkung \& CX-Gatter}

Mathematisch beschreiben wir ein Zwei-QuBit-System durch einen Zustand im Tensorprodukthilbertraum der einzelnen Hilberträume. 
Wir behandeln die mathematischen Einzelheiten hier nicht, sondern konzentrieren uns auf den qualitativen Unterschied zwischen QuBits und Bits, der durch die Möglichkeit der Verschränkung hervorgerufen wird.
Zunächst definieren wir für unseren Zweck einen allgemeinen Zwei-QuBit-Zustand $\ket{q_1q_2}$ als:
\begin{align*}
\ket{q_1q_2} = \begin{pmatrix}
\alpha \\
\beta \\
\gamma \\
\delta
\end{pmatrix} \qquad \alpha, \beta, \gamma, \delta \in \mathbb{R} ,
\end{align*}

mit $\alpha^2 + \beta^2 + \gamma^2 + \delta^2 = 1$. Dabei wählen wir als Basis:
\begin{align*}
\ket{00} = \begin{pmatrix}
1\\
0 \\
0 \\
0
\end{pmatrix},
\ket{01} = \begin{pmatrix}
0\\
1 \\
0 \\
0
\end{pmatrix},
\ket{10} = \begin{pmatrix}
0\\
0 \\
1 \\
0
\end{pmatrix},
\ket{11} = \begin{pmatrix}
0\\
0 \\
0 \\
1
\end{pmatrix} .
\end{align*}

Es seien zwei Ein-QuBit-Zustände gegeben. Dann kann man aus diesen mit dem Tensorprodukt $\otimes$ einen Zwei-QuBit-Zustand bilden:

\begin{align*}
\ket{q_1q_2} :=
\ket{q_1} \otimes \ket{q_2} = \begin{pmatrix}
\alpha_1\\
\alpha_2\\
\end{pmatrix} \otimes \begin{pmatrix}
\beta_1\\
\beta_2\\
\end{pmatrix}
 = \begin{pmatrix}
\alpha_1 \beta_1\\
\alpha_1 \beta_2\\
\alpha_2 \beta_1\\
\alpha_2 \beta_2
\end{pmatrix}.
\end{align*}

Eine Besonderheit der Quanteninformation ist nun, dass es Zwei-QuBit-Zustände gibt, die sich nicht als Tensorprodukt von zwei Ein-QuBit-Zuständen darstellen lassen.
Wir nennen diese Zustände verschränkt. 
Das ist grundlegend verschieden zur Boolschen Algebra, wo wir die Information, die in zwei Bits gespeichert ist, durch zwei einzelne Bits darstellen können. 
Ein bekanntes Beispiel für einen verschränkten Zustand ist der sogenannte Bell-Zustand: 
\begin{align*}
\ket{Bell} := \frac{1}{\sqrt{2}} (\ket{00} + \ket{11}) =  \frac{1}{\sqrt{2}} \begin{pmatrix}
1\\
0\\
0\\
1
\end{pmatrix}.
\end{align*}
Im Unterricht kann ein Koeffizientenvergleich zum Beweis gerechnet werden, zum Beispiel analog zu den Übungen zu Kapitel 8 in~\cite{filk2019}.\\

Wir betrachten nun ein Gatter, das einen Zustand, der sich als Produkt von zwei Ein-QuBit-Zuständen schreiben lässt, in einen verschränkten Zustand überführt.
Wir geben die Wahrheitstabelle des CX-Gatters in Tabelle~\ref{table:CX-Gatter} an.
Dabei steht CX für \textit{controlled} X und bedeutet, dass auf dem zweiten QuBit ein X nur dann ausgeführt wird, wenn das erste QuBit im Zustand $\ket{1}$ ist.
\begin{table}[htb]
\setlength{\tabcolsep}{3.5pt}
\[
\begin{array}{c | c | c | c}
    \ket{q_0} & \ket{q_1} & CX \ket{q_0} & CX \ket{q_1}  \\ \hline 
     \ket{0} & \ket{0} & \ket{0} & \ket{0} \\
     \ket{0} & \ket{1} & \ket{0} & \ket{1} \\
     \ket{1} & \ket{0} & \ket{1} & \ket{1} \\
     \ket{1} & \ket{1} & \ket{1} & \ket{0} \\
\end{array} 
\]
\caption{Wahrheitstabelle des Zwei-QuBit-Gatters CX}\label{table:CX-Gatter}
\end{table}

Das CX-Gatter angewandt auf den Zustand $\ket{+} \otimes \ket{0} = \frac{1}{\sqrt{2}} (\ket{00} + \ket{10})$ ergibt den $\ket{Bell}$-Zustand.
Dabei beschreibt das Gatter eine Spiegelung im vierdimensionalen Raum der Zwei-QuBit-Zustände.
Zwar gibt es für den $\ket{Bell}$-Zustand keine Veranschaulichung durch zwei Einheitskreise, was allerdings schön verdeutlicht, dass ein Zwei-QuBit-Zustand \textit{mehr} ist als zwei Ein-QuBit-Zustände sind.
Was wir nicht zeigen ist, dass man mit dem $CX$-Gatter und den Ein-QuBit-Gattern alle Quantenschaltkreise für eine beliebige Anzahl an QuBits realisieren kann. 

\subsection{Quantenteleportation}

\begin{figure}[H]
\centerline{
\Qcircuit @C=1.3em @R=1.2em {
&  &   &    & &  & & & &   \mbox{ Alice}   \\
   \lstick{\ket{\psi} \rule{0.5em}{0em}}   &   \ctrl{1}   &   \gate{H}   &   \measure{M}   &   \cw  & \cw \cwx[1]   &                           &   &     &     \\
   \lstick{\ket{q^{Alice}_{Bell}} \rule{0.5em}{0em}}  &   \targ      &    \qw    &   \measure{M}   &   \cw \cwx[1]   &   \cwx[1]& & & \push{\rule{0em}{1.5em}} & \\
 &    &   &    &   \cwx[1]       &   \cwx[1]       &   &  &     &   \\
                                           &              &              &                  &   \cwx[1]       &   \cwx[1]       &                           &                           &                                &   \mbox{ Bob}     \\
                                           &              &              &                  &   \cwx[1]       &                 &   \cw                     &   \control \cw  \qwx[2]   & \cw                            &                  \\
                                           &              &              &              &                 &   \cw           &   \control \cw  \qwx[1]   &   \cw                     & \cw                            &                  \\
    \lstick{\ket{q^{Bob}_{Bell}} \rule{0.5em}{0em}}       &   \qw        &   \qw        &   \qw            &   \qw           &   \qw           &   \gate{X}                   &   \gate{Z}                & \push{\rule{0em}{1.5em}} \qw   &   \ket{\psi}
\gategroup{1}{1}{3}{9}{.7em}{.}
\gategroup{5}{1}{8}{9}{.7em}{.}}}
\caption{Quantenteleportation. Schaltkreis für das Quantenteleportationsprotokoll. Alice und Bob haben jeweils ein QuBit eines $\ket{Bell}$-Zustands. Alice führt dann ein $CX$-Gatter und ein $H$-Gatter auf dem QuBit aus, das sie teleportieren möchte und sendet die klassische Information des Messergebnisses an Bob. Dieser wendet, je nach Bitkombination von Alice, null bis zwei Gatter an. So erhält Bob in jedem Fall das QuBit im Zustand $\ket{\psi}$, wie eine Rechnung im Anhang zeigt. Die doppelten Linien verdeutlichen klassische Informationskanäle und einfache Linien QuBit-Kanäle.}\label{fig:quantenteleportation}
\end{figure}
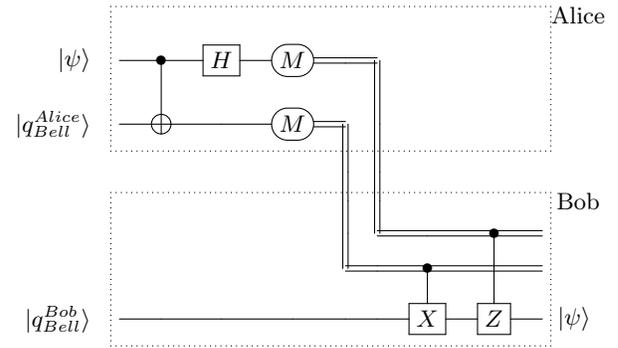

Ziel der Quantenteleportation ist es, die Information eines QuBits zu übertragen, ohne es physikalisch zu versenden.
Dazu benötigen die beiden Kommunikationspartner Alice und Bob jeweils ein QuBit eines Paares, das gemeinsam den Quantenzustand $\ket{Bell}$ beschreibt. 
Wir nennen diese QuBits $\ket{q_{Bell}^{Alice}}$ und $\ket{q_{Bell}^{Bob}}$.
Außerdem werden während des Verfahrens zwei klassische Bits versendet.
Der Schaltkreis in Abbildung~\ref{fig:quantenteleportation} beschreibt die Quantenteleportation. Die ausführlichen Rechnungen für den Schaltkreis finden sich in jedem Standardwerk, zum Beispiel in~\cite{Nielsen_Chuang2011}.

Mit dem besprochenen Verhalten bei einer Messung kann der Schaltkreis der Quantenteleportation nachvollzogen werden. 
Wir geben hier lediglich den Schaltkreis an, die Zwischenergebnisse werden im Anhang, siehe Kapitel~\ref{sec:anhang}, aufgeführt.
Während der Unterrichtseinheit wurden die Zustände nach jedem Gatter berechnet. 
Um eine Tensorschreibweise zu vermeiden, empfehlen wir, die QuBit-Zustände übereinander zu schreiben.
Dadurch kann ikonisch plausibel gemacht werden, wie die Messung der QuBits, die zu Beginn einen Bell-Zustand beschreiben, den Zustand projiziert.
Für den Zustand der drei QuBits vor der Benutzung des Schaltkreises mit $\ket{\psi} = \alpha \ket{0} + \beta \ket{1}$ schreibt man beispielsweise: \\
\begin{align*}
\begingroup
\renewcommand*{\arraystretch}{1.4}
 \ket{ \begin{matrix}
\psi \\
q^{Alice}_{Bell}\\
q^{Bob}_{Bell}\\
\end{matrix}}
\endgroup
=
\frac{\alpha}{\sqrt{2}} \begin{matrix}
\ket{0} \\
\ket{0}\\
\ket{0}\\
\end{matrix}
+
\frac{\beta}{\sqrt{2}} \begin{matrix}
\ket{1} \\
\ket{0}\\
\ket{0}\\
\end{matrix}
+
\frac{\alpha}{\sqrt{2}}  \begin{matrix}
\ket{0} \\
\ket{1}\\
\ket{1}\\
\end{matrix}
+
\frac{\beta}{\sqrt{2}} \begin{matrix}
\ket{1} \\
\ket{1}\\
\ket{1}\\
\end{matrix} .
\end{align*}

Wichtig ist, dass die Messung von zwei QuBits im Allgemeinen nur die ersten beiden QuBits festlegt. 
Das dritte QuBit befindet sich weiterhin in einer Superposition (Überlagerung) der Basiszustände, bis es gemessen wird. 
Das Protokoll der Quantenteleportation kann genutzt werden, um Quanteninformation in einem Netzwerk zu verteilen.
Solche Quantennetzwerke werden aktuell gebaut~\footnote{Siehe Webseite der Quantum Internet Alliance: \url{https://quantum-internet.team/}, zuletzt abgerufen am \today}.
\subsection{Hardware und Physik}
QuBits können durch verschiedene physikalische Systeme realisiert werden.
Da wir die Thematik aus einer informatischen Perspektive behandeln, gehen wir hier nicht weiter auf die physikalischen Details ein und nennen lediglich drei Beispiele:

\begin{enumerate}
\item QuBits können im Polarisationszustand von Photonen codiert werden. Eine schöne physikalische Anwendung ist der Quanten-Zeno-Effekt~\cite{misra1977} beziehungsweise dessen umgekehrte optische Version~\cite{filk2019}.
\item QuBits können durch quantisierte Ströme in supraleitenden Schaltkreisen realisiert werden~\cite{devoret2004}. Diese Realisierung ist besonders für die technische Entwicklung von Quantencomputern relevant.
\item QuBits können als Zwei-Zustandssystem von Atomen oder Ionen implementiert werden~\cite{bruzewicz2019}.
\end{enumerate}

%%%%%%%%%%%%%%%%%%%%%%%%%%%%%%%%%%%%%%%%%%%%%%%%%%%%%
\section{Didaktischer Mehrwert \& Conclusio}\label{sec:conclusion}
%%%%%%%%%%%%%%%%%%%%%%%%%%%%%%%%%%%%%%%%%%%%%%%%%%%%%

Es wurde gezeigt, dass eine Einbettung der Quanteninformationsverarbeitung in den Informatikunterricht der Oberstufe möglich ist.
Die mathematischen Grundlagen sind vorhanden und das Thema findet viel mediale Beachtung, was zu einer hohen Motivation der Schüler führt.\\

Darüber hinaus möchten wir für die Sinnhaftigkeit des Inhalts im Informatikunterricht argumentieren.
Die binäre Diskretisierung von Information ermöglichte der Informatik große Fortschritte.
Sie ist allerdings weniger selbstverständlich, als wir es heute lehren; mit Hilfe der Quanteninformation ist es möglich, einen fundamental unterschiedlichen Informationsträger vorzustellen. 
Dabei können die Vor- und Nachteile gegenüber der herkömmlichen Bits in wenigen Doppelstunden herausgearbeitet werden.
Außerdem ist die Quanteninformationsverarbeitung ein gutes Bindeglied zwischen der Vektoranalysis der Oberstufe und der Gatter- und Wahrheitstabellenschreibweise der Informatik.\\

Offen bleibt die Frage über die Ausbildung der Lehrkräfte in der Quanteninformationsverarbeitung. 
Diese ist derzeit nicht Bestandteil des klassischen Informatiklehramtstudiums, wodurch lediglich eine interessierte und fortgebildete Teilmenge der Lehrerschaft für den Unterricht in Betracht kommt.
Hier kann es sich lohnen, fächerübergreifend mit Physiklehrern zusammenzuarbeiten, da deren Ausbildung die Grundlagen der Quanteninformation in einigen Fällen abdeckt.
In früheren Arbeiten zur Didaktik der Quanteninformationsverarbeitung wurde bereits die Möglichkeit zur Bearbeitung des Stoffes in der Schule thematisiert~\cite{dur2010, dur2012, kohnle2016}.
Wir sind der Überzeugung, dass die Einbettung der Quanteninformationsverarbeitung sowohl im Physik- als auch im Informatikunterricht aus unterschiedlichen Gründen - im Fall der Informatik aus den genannten - sinnvoll ist.
Die vorliegende Arbeit ist dabei nur ein erster Schritt in diese Richtung. 
Es benötigt weitere Anstrengungen in der didaktischen Aufarbeitung der Thematik sowie der Erstellung von Unterrichtsmaterialien.

%%%%%%%%%%%%%%%%%%%%%%%%%%%%%%%%%%%%%%%%%%%%%%%%%%%%%
\section{Danksagung}\label{sec:danksagung}
%%%%%%%%%%%%%%%%%%%%%%%%%%%%%%%%%%%%%%%%%%%%%%%%%%%%%
An erster Stelle gilt ein besonderer Dank Urs Lautebach: für die Motivation, den Versuch dieser Einheit zu wagen und für die Einbettung der Einheit in seinen Unterricht. 
Darüber hinaus danke ich 
Thomas Filk,
Clara Fuchs
und
Franziska Gerke 
für den hilfreichen fachlichen Austausch, Jan Labusga danke ich für einen strengen Blick auf die Rechtschreibung.
Meinem Doktorvater Andreas Buchleitner danke ich für die kontinuierliche Unterstützung meiner Arbeit und die Förderung meiner Interessen.
Abschließend danke ich der Konrad-Adenauer-Stiftung für die finanzielle Unterstützung dieser Arbeit.

%%%%%%%%%%%%%%%%%%%%%%%%%%%%%%%%%%%%%%%%%%%%%%%%%%%%%
\section{Anhang}\label{sec:anhang}
%%%%%%%%%%%%%%%%%%%%%%%%%%%%%%%%%%%%%%%%%%%%%%%%%%%%%

\begin{figure}[H]
\centerline{
\Qcircuit @C=1.3em @R=1.2em {
&  &   &    & &  & & & &   \mbox{ Alice}   \\
   \lstick{\ket{\psi} \rule{0.5em}{0em}}   &   \ctrl{1}   &   \gate{H}   &   \measure{M}   &   \cw  & \cw \cwx[1]   &                           &   &     &     \\
   \lstick{\ket{q^{Alice}_{Bell}} \rule{0.5em}{0em}}  &   \targ      &    \qw   \ar@{.}[]+<1.5em,6em>;[d]+<1.5em,-8em>    \ar@{.}[]+<-1.5em,6em>;[d]+<-1.5em,-8em>   &   \measure{M}   &   \cw \cwx[1]   &   \cwx[1]& & & \push{\rule{0em}{1.5em}} & \\
 &    &   &    &   \cwx[1]       &   \cwx[1]       &   &  &     &   \\
                                           &              &              &                  &   \cwx[1]       &   \cwx[1]       &                           &                           &                                &   \mbox{ Bob}     \\
                                           &              &              &                  &   \cwx[1]       &                 &   \cw                     &   \control \cw  \qwx[2]   & \cw                            &                  \\
                                           &              &              &              &                 &   \cw           &   \control \cw  \qwx[1]   &   \cw                     & \cw                            &                  \\
    \lstick{\ket{q^{Bob}_{Bell}} \rule{0.5em}{0em}}       &   \qw        &   \qw        &   \qw            &   \qw           &   \qw           &   \gate{X}                   &   \gate{Z}                & \push{\rule{0em}{1.5em}} \qw   &   \ket{\psi} \\
\lstick{\ket{\phi_1} \rule{0.5em}{0em}} &  \ket{\phi_2} &   \ket{\phi_3} &    & &  & & & &  \ket{\psi}
\gategroup{1}{1}{3}{9}{.7em}{.}
\gategroup{5}{1}{8}{9}{.7em}{.}}}
\caption{Quantenteleportation. Schaltkreis für das Quantenteleportationsprotokoll mit Markierungen für die jeweiligen Zustände, die im Anhang angegeben sind.}\label{fig:quantenteleportationausfuehrlich}
\end{figure}
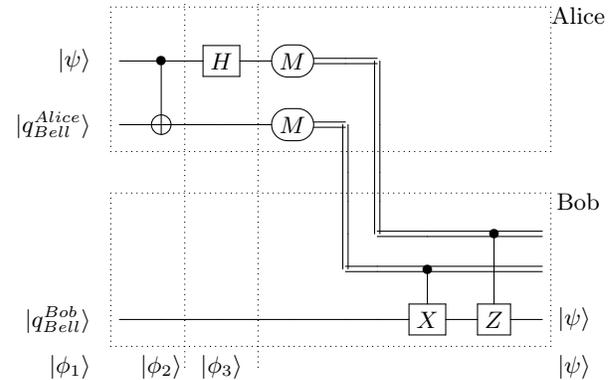

In Abbildung~\ref{fig:quantenteleportationausfuehrlich} wird erneut der Schaltkreis für die Quantenteleportation gezeigt. 
Wir geben hier die drei Zustände: $\ket{\phi_1}, \ket{\phi_2}, \ket{\phi_3}$ zu verschiedenen Zeitpunkten des Schaltkreises an:
\begin{align*}
\ket{\phi_1} = 
\begingroup
\renewcommand*{\arraystretch}{1.4}
 \ket{ \begin{matrix}
\psi \\
q^{Alice}_{Bell}\\
q^{Bob}_{Bell}\\
\end{matrix}}
\endgroup
&=
\frac{\alpha}{\sqrt{2}} \begin{matrix}
\ket{0} \\
\ket{0}\\
\ket{0}\\
\end{matrix}
+
\frac{\beta}{\sqrt{2}} \begin{matrix}
\ket{1} \\
\ket{0}\\
\ket{0}\\
\end{matrix}
+
\frac{\alpha}{\sqrt{2}}  \begin{matrix}
\ket{0} \\
\ket{1}\\
\ket{1}\\
\end{matrix}
+
\frac{\beta}{\sqrt{2}} \begin{matrix}
\ket{1} \\
\ket{1}\\
\ket{1}\\
\end{matrix} 
\\
\ket{\phi_2} = 
\begingroup
\renewcommand*{\arraystretch}{1.4}
 \ket{ \begin{matrix}
\psi \\
q^{Alice}_{Bell}\\
q^{Bob}_{Bell}\\
\end{matrix}}
\endgroup
&=
\frac{\alpha}{\sqrt{2}} \begin{matrix}
\ket{0} \\
\ket{0}\\
\ket{0}\\
\end{matrix}
+
\frac{\beta}{\sqrt{2}} \begin{matrix}
\ket{1} \\
\ket{1}\\
\ket{0}\\
\end{matrix}
+
\frac{\alpha}{\sqrt{2}}  \begin{matrix}
\ket{0} \\
\ket{1}\\
\ket{1}\\
\end{matrix}
+
\frac{\beta}{\sqrt{2}} \begin{matrix}
\ket{1} \\
\ket{0}\\
\ket{1}\\
\end{matrix} \\
\ket{\phi_3} = 
\begingroup
\renewcommand*{\arraystretch}{1.4}
 \ket{ \begin{matrix}
\psi \\
q^{Alice}_{Bell}\\
q^{Bob}_{Bell}\\
\end{matrix}}
\endgroup
&=
\frac{\alpha}{\sqrt{2}} \begin{matrix}
\ket{0} \\
\ket{0}\\
\ket{0}\\
\end{matrix}
+
\frac{\alpha}{\sqrt{2}} \begin{matrix}
\ket{1} \\
\ket{0}\\
\ket{0}\\
\end{matrix}
+
\frac{\beta}{\sqrt{2}}  \begin{matrix}
\ket{0} \\
\ket{1}\\
\ket{0}\\
\end{matrix}
-
\frac{\beta}{\sqrt{2}} \begin{matrix}
\ket{1} \\
\ket{1}\\
\ket{0}\\
\end{matrix}\\
&+ \frac{\alpha}{\sqrt{2}} \begin{matrix}
\ket{0} \\
\ket{1}\\
\ket{1}\\
\end{matrix}
+
\frac{\alpha}{\sqrt{2}} \begin{matrix}
\ket{1} \\
\ket{1}\\
\ket{1}\\
\end{matrix}
+
\frac{\beta}{\sqrt{2}}  \begin{matrix}
\ket{0} \\
\ket{0}\\
\ket{1}\\
\end{matrix}
-
\frac{\beta}{\sqrt{2}} \begin{matrix}
\ket{1} \\
\ket{0}\\
\ket{1}\\
\end{matrix} .
\end{align*}

Man erkennt nun durch Umsortieren der Summanden von Zustand $\ket{\psi_3}$, dass nach Messung von QuBit eins und zwei ein Überlagerungszustand auf QuBit drei vorliegt, der dem zu teleportierenden QuBit $\ket{\psi} = \frac{\alpha}{\sqrt{2}} \ket{0} + \frac{\beta}{\sqrt{2}} \ket{1}$ bis teilweise auf Vorzeichen und Vertauschung der Vorfaktoren $\alpha, \beta$ gleicht.
Zum Beispiel im Falle einer Messung von 1 für QuBit eins und 0 für QuBit zwei, liegt der Zustand $\frac{\alpha}{\sqrt{2}} \ket{0} - \frac{\beta}{\sqrt{2}} \ket{1}$ vor, der durch ein $Z$-Gatter in den Ausgangszustand des zu teleportierenden QuBits $\ket{\psi}$ überführt wird.
Mit den bedingten Gattern wird so in jedem der vier Fälle dafür gesorgt, dass Bob den Zustand $\ket{\psi}$ erhält. \\

Interessanterweise ist hierbei anzumerken, dass Bobs QuBit nach der Standarddeutung der Quantenmechanik instantan in den entsprechenden Zustand springt, sobald die Messung von Alice verursacht wird.
Allerdings muss Alice klassische Information an Bob senden, damit er die bedingten Gatter anwenden kann. 
Ansonsten, so kann man zeigen, kann er keine Information über Alices QuBit gewinnen.
Diese Einschränkung führt dazu, dass in diesem Protokoll zwar ein Quantenzustand teleportiert wird, allerdings Information trotzdem nur mit endlicher Geschwindigkeit übertragen wird.

\bibliography{QC}
\bibliographystyle{unsrt}

\end{document}